\documentclass[12pt]{iopart}
\usepackage{graphicx}
\usepackage{color}
\begin{document}

\title[Spatial superposition using a levitated ferrimagnetic nanoparticle]{\textbf{Large spatial Schrodinger cat state using a levitated ferrimagnetic nanoparticle}}

\author{A. T. M. Anishur Rahman}
\ead{a.rahman@ucl.ac.uk}
\address{Department of Physics and Astronomy\\
	University College London\\
	Gower Street, WC1E 6BT
	London, UK} 
%
%\author{Content \& Services Team}
%
%\address{IOP Publishing, Temple Circus, Temple Way, Bristol BS1 6HG, UK}
%\ead{submissions@iop.org}
%\vspace{10pt}
%\begin{indented}
%\item[]August 2017
%\end{indented}

\begin{abstract}
The superposition principle is one of the main tenets of quantum mechanics. Despite its counter-intuitiveness, it has been experimentally verified using electrons, photons, atoms, and molecules. However, a similar experimental demonstration using a nano or a micro particle is non-existent. Here in this article, exploiting macroscopic quantum coherence and quantum tunneling, we propose an experiment using a  levitated magnetic nanoparticle to demonstrate such an effect. It is shown that the spatial separation between the delocalized wavepackets of a $20~$nm ferrimagnetic yttrium iron garnet (YIG) nanoparticle can be as large as $5~$$\mu$m. We argue that, in addition to using for testing one of the most fundamental aspects of quantum mechanics, this scheme can simultaneously be used to test different modifications, such as wavefunction collapse models, to the standard quantum mechanics. Furthermore, we show that the spatial superposition of a core-shell structure, a YIG core and a non-magnetic silica shell, can be used to probe quantum gravity.
\end{abstract}

%\maketitle

\section{Introduction} 
Quantum mechanics permits an object, however big, to be spatially delocalized in two different places at once \cite{Monroe1996,Arndt1999,Eibenberger2013,Arndt2014}. Despite being counter-intuitive and in direct conflict with our everyday experience, the superposition principle has been experimentally verified using neutrons \cite{ZawiskyM2002}, electrons \cite{Arndt2014}, ions \cite{Monroe1996} and molecules \cite{Arndt1999,Eibenberger2013}. The current record for the largest spatial superposition is $0.5~$m which was realized using a Bose-Einstein condensate of Rubidium atoms in an atomic fountain \cite{Kovachy2015}, while the heaviest object so far put into a superposition state is about $1\times 10^{-23}~$ kg \cite{Eibenberger2013}. However, a similar test using a mesoscopic ($\approx 100~$nm) object is still missing and it is one of the most pursued problems in modern quantum mechanics \cite{Marshall2003,RomeroPRL2012,Zhang2013,Arndt2014,BatemanNatComm2014,WanPRL2016,Romero_IsartNJP2017,FrowisRevModPhys2018,RahmanATMAnishur2018}. A successful demonstration of such a state can testify various modifications to the quantum mechanics e.g. wavefunction collapse models \cite{Bassi2013,FrowisRevModPhys2018}, decoherence mechanisms such as gravitational state reduction \cite{Penrose1996OGri}, measurement hypothesis \cite{Arndt2014} and the apparent conflict between relativity and quantum mechanics \cite{Zurek2003,Arndt2014}. Furthermore, apart from being of pure fundamental interest, a macroscopic superposition state is also of significant practical relevance due to the emergence of quantum technologies e.g. quantum computing and communications \cite{Bennet2000}. That is the superposition principle is the essential ingredient of quantum computing \cite{Bennet2000} as well as behind the absolute security of quantum communications \cite{QuantumMiller2008}. Understanding the superposition principle at the macroscopic level can enrich our knowledge about the nature around us and can improve metrology, and quantum computing and communications \cite{FrowisRevModPhys2018}.

In this article, we propose an experimental scheme for creating a spatial superposition state by exploiting the superposition that naturally occurs when two potential wells are coupled together with a potential barrier in between them. In particular, due to tunneling, in magnetically ordered material such as ferromagnet and ferrimagnet with magnetocrystalline anisotropy, degeneracy among different spin states are lifted \cite{AwschalomScience1992,Garg1994,Chudnovsky_tejada_1998,TejadaNanothech2001,HillScience2003} (see Fig. \ref{fig1}). In these systems the ground state is the symmetric superposition of all-up and all-down spin states \cite{ChudnovskyPRL2000,Chudnovsky_tejada_1998,TejadaNanothech2001}. Exploiting this naturally occurring spin superposition, and a magnetic field gradient, we propose a scheme for creating a spatial Schrodinger cat state. We show that the separation between the delocalized superposed states is significantly larger than the object involved in the superposition and indeed can be as large as $5~\mu$m. The mass of this object is $2\times 10^{-20}~$kg.

Note that macroscopic quantum coherence (MQC), coherent evolution of many spins - a key requirement for the current proposal, has been studied extensively in the past- both theoretically \cite{ChioleroPRB1997,ChioleroPLR1998,Garg1994,Chudnovsky_tejada_1998,ChudnovskyPRL2000} and experimentally \cite{AwschalomPRL1992,AwschalomScience1992,GiderScience1995,delBarcoEPL1999,delBarcoPRB2000,SchlegelPRL2008}. For example, MQC has been experimentally confirmed in molecular magnets consisting of manganese clusters \cite{HillScience2003} with $S=9$ and iron based system \cite{delBarcoEPL1999,delBarcoPRB2000} with $S=10$. Similarly, quantum coherence has been demonstrated in nanomagnets e.g. ferritin- a naturally occurring protein about $7.5~$nm in diameter with an antiferromagnetic core and uncompensated spins \cite{AwschalomPRL1992,AwschalomScience1992,GiderScience1995}. In this case the number of spin involved in the coherence experiment was $\approx 300$ or $S=150$.

\section{Spatial superposition} 
\begin{figure}
	\centering
	\includegraphics[width=15cm]{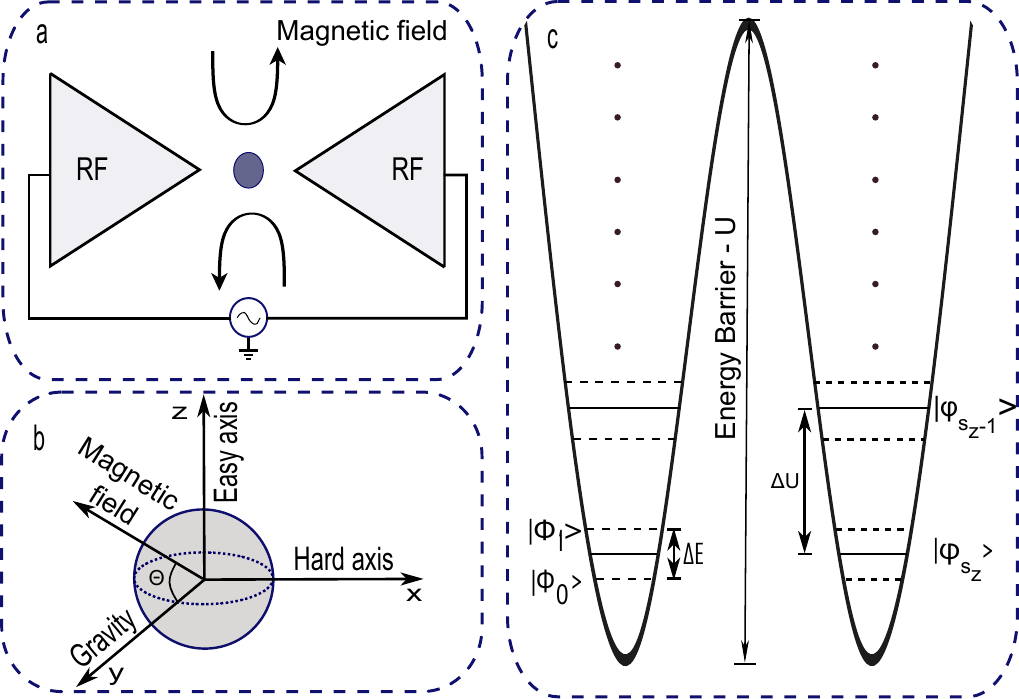}
	\caption{\label{fig1} Experimental schematics - a) Ion trap including magnetic field, b) A yttrium iron garnet (YIG) nanocrystal with its easy axis align to the $z$-axis. Earth's gravity points along the $y-$axis. c) Double potential well. Solid lines represent spin states when only one potential well is present while dashed lines show spin states when two wells are coupled. $\Delta E$ is the energy gap between the ground state $|\phi_0\rangle$ and the first excited state $|\phi_1\rangle$ when two potential wells are coupled while $\Delta U$ is the same difference in energy when only one potential well is considered.}
\end{figure}

A schematic of the proposed experiment is shown in Fig. \ref{fig1}a. In this scheme a single domain magnetic nanoparticle of radius $R$, volume $V$, mass $m$, spin $\mathbf{S}$ and its easy axis aligned to $z-$axis or the quantization axis (see Fig. \ref{fig1}b) is levitated using an ion trap \cite{AldaAPL2016,Huillery2019} at a cryogenic temperature ($\approx 300~$mK \cite{VinantePRA2019}). After levitation, the centre-of-mass (CM) temperature $T_{cm}$ of the particle is reduced to mK level using parametric feedback cooling \cite{AldaAPL2016}. Here, one can use a superconducting quantum interference device (SQUID) for the detection and the manipulation of the CM motion of the levitated particle \cite{VinantePRL2017,JohnssonSciRep2016}. Furthermore, we assume that $S$ is an integer to ensure that tunneling between two wells, discussed below, is permissible \cite{LossPRL1992}. Additionally, we will show that tunneling remains valid when one considers the physical rotation of the nanoparticle that may arise when spins tunnel from one well to the other \cite{ChudnovskyPRB2010}. 

%\begin{center}
%\end{center}

In a single domain ferromagnet, antiferromagnet and ferrimagnet, all spins are aligned and coupled together due to exchange interaction \cite{AwschalomPRL1992,AwschalomScience1992,GiderScience1995,WernsdorferPRL1997,ChioleroPRB1997,ChudnovskyPRL2000}. The exchange interaction can be represented as $-\sum_{i\ne j}J\mathbf{s}_i.\mathbf{s}_j$, where $J$ is the strength of the exchange coupling (for YIG $J\approx 7~$meV \cite{PrincepQMaterials2017}), and $\textbf{s}_i$ and $\textbf{s}_j$ are the spin of the neighbouring $i$th and $j$th atoms. Furthermore, due to magnetocrystalline anisotropy, there is a certain direction inside the crystal along which spins are preferentially aligned (easy axes, $z-$axis, see Fig. \ref{fig1}b) \cite{GiderScience1995,ChioleroPRB1997,Chudnovsky_tejada_1998}. Under this condition, spin $\mathbf{S}$ can have two opposite orientations, $|S_z\rangle$ and $|-S_z\rangle$, of equal energy along the easy axis separated by an energy barrier $U=K_iV=-DS_z^2$ with $K_i=K_x$, $K_y$ and $K_x>>K_y>0$, where $K$'s and $D$ are the magnetocrystalline anisotropy constants. Equivalently, due to the presence of magnetocrystalline anisotropy, there exists two potential wells in which the orientation of the spins are opposite (Fig. \ref{fig1}c). In isolation, each of these potential wells contains $S$ spin levels $|\psi_{m}\rangle$ with $m=\pm 1, \pm 2...\pm S_z$. The separation in energy between two such consecutive spin states in a well is $\Delta U = D(2m-1)$. Energetically, spin levels in the two isolated wells with the same $|m|$ values are equal or the states are degenerate. However, due to the coupled nature of the potential wells degeneracy is lifted and the eigenstates of the overall system \cite{TejadaNanothech2001,Leuenberger2001} are now the symmetric and antisymmetric superposition of the eigenstates of the individual well e.g. $|\phi_n\rangle=(|\psi_m\rangle \pm |\psi_{-m}\rangle)/\sqrt{2}$, where $n=0, 1, 2...2S_z-1$, and $m=1,2,3,....S_z$. The ground state of this system is $|\phi_0\rangle=(|\psi_{S_z}\rangle+|\psi_{-S_z}\rangle)/\sqrt{2}$ while the first excited state is $|\phi_1\rangle=(|\psi_{S_z}\rangle-|\psi_{-S_z}\rangle)/\sqrt{2}$. The separation in energy between the ground state and the first excited state or the so-called tunnel splitting \cite{ChioleroPRB1997} is given by $\Delta E=\hbar\omega_0\exp{(-S\sqrt{K_y/K_x})}$, where $\hbar$ is the reduced Planck constant and $\omega_0\approx 10^{11}-10^{13}$ Hz is the characteristic frequency \cite{Garg1994,ChioleroPRB1997}. Depending on the material under consideration, $\Delta E$ can be several hundred millikelvin while $\Delta U$ can be tens of kelvin \cite{TejadaNanothech2001}. $\Delta E$ can be controlled by applying a weak magnetic field orthogonal to the crystal's easy axis and hence can be tuned \cite{delBarcoEPL1999,delBarcoPRB2000,TejadaNanothech2001}. In contrast, a magnetic field along the easy axis of the magnetic nanoparticle lifts the degeneracy and as the degeneracy is removed tunneling disappears along with it \cite{TejadaNanothech2001}. One can exploit this feature as a control mechanism to initialize or remove a spin superposition as required. Indeed, in the proposed experiment, a weak d.c. magnetic field $B_0$ is activated whenever a magnetic particle is trapped. This confines the spins in one of the wells and aligns the particle's easy axis along the direction of the magnetic field. This magnetic field and the low temperature considered here forces the overall system to either $|\psi_{S_z}\rangle$ or $|\psi_{-S_z}\rangle$ state.

After the initial state preparation such as attaining the desired CM and internal temperatures, magnetic field $B_0$ is switched off. This initiates tunneling and hence a spin superposition. Given the low experimental temperature ($300~$mK) and the relevant tunnel splitting $\Delta E \approx 500~$mK (see below), population in all states except $|\phi_0\rangle=(|\psi_{S_z}\rangle+|\psi_{-S_z}\rangle)/\sqrt{2}$ can be safely ignored. We use $|\phi_0\rangle$ for the creation of a spatial Schrodinger cat. At this stage the ion trap is switched off and an inhomogeneous magnetic field is activated \cite{Harrison2015}. The direction of the magnetic field gradient is such that it makes an angle $\theta$ with the direction of the earth's gravity (along $y-$axis, Fig. \ref{fig1}b). The untrapped particle evolves under the influence of gravitational and magnetic fields for a suitable time $t$. At this state the Hamiltonian is \cite{WanPRL2016}

\begin{eqnarray}
\hat{H}=\frac{\hat{p_0}^2}{2m}-g_L\mu_B\frac{dB}{dz}\hat{S_z}\hat{z}+mg\cos{\theta}\hat{y},
\end{eqnarray}

where $m$ is the mass of the levitated particle, $\mu_B$ is the Bohr magneton, $dB/dz$ is the magnetic field gradient, $g_L$ is the Lande factor and $g$ is the gravitational acceleration. $\hat{p_0}$ is the momentum before the particle was released from the trap. At time $t_0/4$ the initial magnetic field gradient is switched off and a new magnetic field gradient of opposite polarity to that of the original magnetic field gradient is activated. This new field gradient redirects wavepackets towards the center. Here, the activation (deactivation) of the magnetic field gradient is carried out by slowly increasing (decreasing) the magnitude of the field in such a way that it does not create a sudden impulse on the nanoparticle. At time $3t_0/4$, the polarity of the field gradient is changed for the last time which decelerates the wavepackets as they approach each other from the opposite directions. Finally, at time $t_0$, the magnetic field gradient is completely switched off. This ensures two wavepackets overlap exactly with each other at the center. At this stage, the ion trap is turned back on to recapture the particle and simultaneously a spin measurement along the $x-$axis is carried out. Here, owing to the different trajectories of the wavepackets through the gravitational field, a gravity induced phase difference $\beta_g=(1/16\hbar)gt_0^3g_LS_z\mu_B(dB/dz)\cos{\theta}$ between the wavepackets is accrued e.g. $|\phi_0\rangle=(|\psi_{S_z}\rangle+e^{-i\beta_g}|\psi_{-S_z}\rangle)/\sqrt{2}$ \cite{WanPRL2016}. The effect of this phase appears in the spin measurement where the probability of measuring $|\pm S_x\rangle$ varies as $1\pm \cos{\beta_g}$. Since spin cannot acquire a phase due to the different trajectories through the gravitational field, any effect of this phase difference on the spin measurement is considered as an evidence of the spatial superposition created \cite{WanPRL2016}. One can use $t_0$ and $\theta$ to give a controllable phase in the spin measurement. To build up statistics, the sequence of events described above can be carried out as many times as required. The maximum spatial separation between the two arms of the superposed states is achieved just before the two wavepackets start approaching each other from the opposite directions and is given by \cite{WanPRL2016}

\begin{eqnarray}
\Delta z =\frac{g_L\mu_B S_z t_0^2}{8m}\frac{dB}{dz},
\label{eq2}
\end{eqnarray}

where $t_0$ is the spin coherence time.

\section{Experiment} Since tunneling is a very general phenomenon in magnetic systems, any magnetic material with a magnetocrystalline anisotropy can be used as a model system for the current proposal. For example, one can use ferritin nanoparticles with $S\approx 150$. With ferritin, macroscopic quantum coherence has already been demonstrated \cite{AwschalomPRL1992,GiderScience1995,ChioleroPRB1997}. Nevertheless, in this article we aim to use yttrium iron garnet (YIG), one of the best known ferrimagnetic materials \cite{CHEREPANOV199381,Serga_2010} with four uncompensated Fe$^{3+}$ ($s=5/2$) atoms per unit cell (lattice constant $a\approx1.5~$ nm) \cite{PrincepQMaterials2017} as a model system. In bulk YIG crystal, spin coherence time ($T_2$) on the order of microseconds has been measured \cite{KaplanPRL1965,KaplanJAP1969,HueblPRL2013}. YIG also relaxes some of the experimental requirements involved. Specifically, YIG is an insulator which ensures no conducting electron and hence no decoherence due to the electric current that a free electron carries. Another advantage of YIG is its high blocking temperature $T_B=64~$K \cite{RAJENDRAN2006} which prevents superparamagnetic behaviour. Furthermore, YIG has the lowest known Gilbert damping $\alpha$ of all known materials \cite{MairFlaigPRB2017}. It determines how a spin system loses energy and angular momentum. In the absence of inhomogeneity, Gilbert damping is related to the spin coherence time $t_0$ via the relation $t_0 =1/\alpha \gamma_r B$ \cite{CapuaAmir2017}, where $\gamma_r$ is the gyromagnetic ratio and $B$ is the magnetic field. $\alpha = 1\times 10^{-5}$ has been measured at $20~$K and according to the theory, in the absence of inhomogeneity - valid for small nanoparticles, it should vanish as the temperature decreases \cite{MairFlaigPRB2017}.

\begin{center}
\begin{figure}
	\includegraphics[width=15cm]{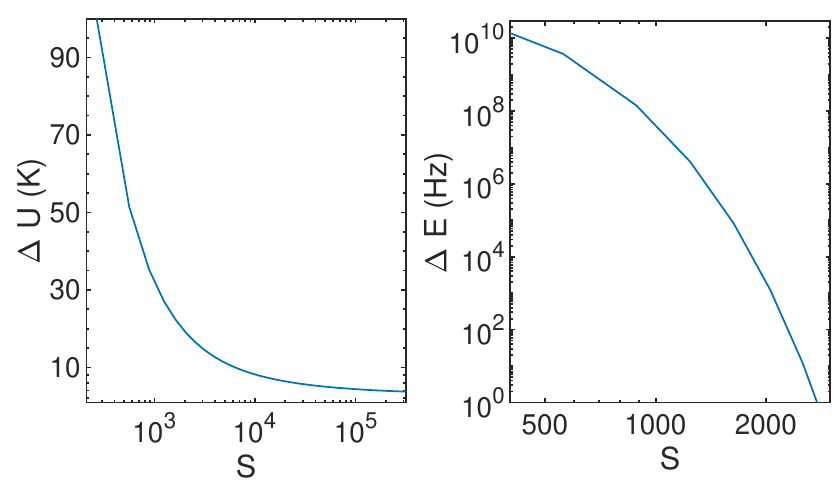}
	\caption{\label{fig2} a) Energy gap $\Delta U$ between $|\psi_{S_z}\rangle$ and $|\psi_{(S_z-1)}\rangle$ as a function of uncompensated spin $S$. Similar results are also valid for $|\psi_{-S_z}\rangle$ and $|\psi_{-(S_z-1)}\rangle$ spin states. b) Tunnel splitting $\Delta E$ or the difference in energy between the ground state $|\phi_0\rangle=(|\psi_{S_z}\rangle+|\psi_{-S_z}\rangle)/\sqrt{2}$ and the first excited state $|\phi_1\rangle=(|\psi_{S_z}\rangle-|\psi_{-S_z}\rangle)/\sqrt{2}$ as a function of $S$.}
\end{figure}
\end{center}

A large spatial separation between the superposed states or a large Schrodinger cat is highly desirable \cite{Wineland2013,Arndt2014} and can be achieved by using a large $S$ (see Eq. (\ref{eq2})). However, a large $S$ accompanies a reduced $\Delta U=DS_z^2$ which ultimately necessitates a lower experimental temperature to avoid excited state $|\phi_{n>1}\rangle$ population. Fig. \ref{fig2}a shows $\Delta U$ as a function of $S$, where we have used $D=K_xV/S_z^2$ \cite{TejadaNanothech2001} and $K_x\approx 5.54\times 10^4~$J~m$^{-3}$ \cite{RAJENDRAN2006}. We have also taken two layers of dead spins on the surface into consideration \cite{RAJENDRAN2006}. It is obvious that $\Delta U$ decreases drastically as $S$ increases. A large $S$ also indicates a reduced tunnel splitting - $\Delta E=\hbar\omega_0\exp{(-S\sqrt{K_y/K_x})}$. To calculate $\Delta E$, one requires $\omega_0$ and $\sqrt{K_y/K_x}$. While the measure of $K_x$ is readily available \cite{RAJENDRAN2006}, experimental values of $K_y$ and $\omega_0$ of YIG nanoparticles can not be found in the literature. However, experiments involving ferritins \cite{AwschalomPRL1992,GiderScience1995,ChioleroPRB1997}, a Fe$^{3+}$ based nanomagnet like YIG, have found $\omega_0/2\pi\approx 10^{12}~$Hz. In Fig. \ref{fig2}b, we have used $\sqrt{K_y/K_x}=10^{-2}$ and $\omega_0/2\pi=10^{12}~$Hz. From Fig. \ref{fig2}b, it is clear that $\Delta E$ reduces severely as $S$ increases. Consequently, one needs to choose $S$ carefully to ensure both $\Delta E$ and $\Delta U$ remain as large as possible. A large $\Delta U$ guarantees, for example, a higher minimum experimental temperature which is beneficial for experiments. Furthermore, a large $S$ can lead to a strong interaction between the system and the environment which can induce rapid decoherence \cite{Chudnovsky_tejada_1998}. For the discussion that follows we take $S=500$ which provides $\Delta U/k_B \approx 50~$K and $\Delta E /h \approx 10~$GHz ($500~$mK) - both of which are experimentally feasible. $S=500$ corresponds to $200$ uncompensated Fe$^{3+}$ atoms and the diameter of the YIG nanoparticle is $\approx 20~$nm. 

It is also instructive to consider the conservation of angular momentum $L$ associated with spin tunneling \cite{ChudnovskyPRB2010}. Specifically, when spins tunnel from one well to the other, to conserve $L$, the particle needs to rotate physically. This may lift the degeneracy unless the rotational energy $L^2/2I$, where $I=\frac{2mR^2}{5}$ is the moment of inertia of a sphere, is dominated by the energy reduced ($\Delta E/2$) due to tunneling \cite{ChudnovskyPRB2010}. In other words, $\alpha=\frac{(\hbar S)^2}{\Delta E I} << 1$, where we have assumed $L=\hbar S$. In our case, for $S=500$ and the mass density of YIG equals to $\rho=5000~$kg$~m^{-3}$, we have $\alpha=5\times 10^{-4}$. This is significantly less than unity and as a result physical rotation of the particle is not expected to have any significant effect on the tunneling.

Finally, let us now consider a numerical example. For that we take $\frac{dB}{dz}=10^6~$T$~$m$^{-1}$ \cite{MaminNatNano2007,Harrison2015} and $t_0=10~\mu$s. On substitution of the relevant values in Eq. (\ref{eq2}), one gets $\Delta z\approx 5~$$\mu$m. This is a macroscopic distance and can be visualized using unaided eyes.

\section{Decoherence} As the macroscopicity of a quantum system increases, so does the possibility of rapid decoherence. Consequently, great care needs to be exercised to avoid this detrimental effect. One such major source of decoherence is the fluctuating magnetic field that may exist around the experiment. However, this can be effectively reduced to picotesla level or $\approx 30$ Hz using a superconducting shield \cite{Hinterberger_2017}. This is significantly lower than the $10~$GHz tunnel splitting found above. Since the proposed experiment is planned to be carried out in a cryogenic condition, adopting a superconducting shield should be relatively straight forward. A further source of decoherence is the nuclear spins \cite{Chudnovsky_tejada_1998,GargPRL1993} which, along with other sources of decoherence e.g. impurities, appears as the linewidth broadening in ferromagnetic resonance (FMR) \cite{SpencerPRL1959,GargPRL1993,Chudnovsky_tejada_1998,MairFlaigPRB2017}. Nevertheless, YIG has the lowest known FMR linewidth of all materials \cite{SpencerPRL1959,HueblPRL2013,MairFlaigPRB2017}. This can be further reduced by eliminating rare-earth contaminants \cite{SpencerPRL1959,GargPRL1993,MairFlaigPRB2017}. For example, by reducing the contents of rare-earth impurities, Spencer et al. \cite{SpencerPRL1959} managed to suppress FMR linewidth by $50$ times. By selectively eliminating $^{57}$Fe atoms from YIG  or by isotropic purification one can improve the coherence time further \cite{GargPRL1993}.  Magnons, collective oscillations of spins in ordered magnetic system e.g. ferrimagnet, can induce decoherence. However, due to the small physical size of the nanoparticle ($R=10~$nm), propagating magnons are irrelevant \cite{Roschmann1977} owing to the high energy excitation $\ge 0.02 cR^{-1}~$Hz involved, where $c$ is the speed of light in free space. To excite magnetostatic modes or the precessional modes \cite{KittelPR1948}, one needs a magnetic field at an angle with the spin quantization axis. Since a superconducting shield will be in use to reduce the background magnetic field ($B_g$) to picotesla level, the effect of these low frequency ($g_L\mu_B B_g/\hbar$) disturbances can be safely ignored. Furthermore, sub-kelvin experimental temperature may be useful in suppressing magnons.

Apart from the decoherence of spins, decoherence of the centre-of-mass motion of the nanoparticle is also of critical importance \cite{Chang2010}. In particular, decoherence of the CM motion can reduce the visibility of the relevant matter-wave interference pattern. However, this can be easily counteracted by performing the experiment in ultra high vacuum ($10^{-9}~$mBar). Incidently, this level of vacuum is readily achievable in cryogenic environment \cite{VinantePRA2019}. Assuming residual helium gas pressure $P=10^{-9}~$mBar, gas temperature $T=300~$mK, helium mass $m_g\approx 6.64\times 10^{-27}~$kg, velocity of the helium atoms $v=\sqrt{k_B T/m_g}\approx 25~$m/s, and the size of nanoparticle $R=10~$nm, the expected number of collisions between the sphere and the gas molecules is $\pi Pv R^2/k_BT\approx 200$ in a second or $2\times 10^{-3}$ collisions during the actual time of the experiment ($10~\mu s$) \cite{Chang2010}. In another word, a collision is very rare. Nevertheless, in the event of an elastic collision with a gas molecule, additional velocity acquired by the YIG particle is $\approx 2\times 10^{-5}~$m/s. This can create a maximum uncertainty of $\approx 0.2~$ nm in the distance traversed by the particle in $10~\mu s$. In contrast, the actual distance travelled by the YIG nanoparticle in the same time is at least $\Delta z=5~\mu m$ or the size of the superposition. This is about four orders of magnitude larger than the uncertainty. Consequently, the effect of a collision between the YIG nanoparticle and a gas molecule on the visibility of the superposition is negligible. Likewise, it can be shown that the decoherence due to the blackbody absorption and emission by the particle is also very small \cite {Chang2010}. Specifically, the amount of power emitted by a nanoparticle of surface area $A$ at temperature $T$ is given by the Stefan-Boltzmann law - $\sigma A T^4$, where $\sigma$ is the Stefan-Boltzmann constant \cite{StillmanGregory}. For the sake of an estimate, let us assume that all the power emitted by the nanoparticle is at the peak emission wavelength $\lambda_{max}=\frac{2.89\times 10^{-3}}{T}$ of the relevant blackbody emission spectrum. Then the number of blackbody photons emitted in a second is $N=2.89\times 10^{-3} \frac{\sigma A T^3}{\hbar c}$. In our case, this is equivalent to $1.85\times 10^{-2}$ photons in a second or $1.85\times 10^{-7}$ photons in $10~\mu s$. In the unlike event of a blackbody photon emission, extra velocity gained by the YIG particle is $\approx 3\times 10^{-12}~$m/s. This will create a position uncertainty of $3\times 10^{-17}~$m - which is vanishingly small. Additionally, it can be shown that the decoherence due to blackbody absorption is also negligible as found by others \cite{Chang2010}. Finally, the effect of vibration associated with the cryogenic environment needs to be accounted. Here, to negate this effect, one can switch off the cryogenics, possible in pulse tube based systems, for the duration of the experiment ($10~\mu s$). Alternatively, one can use a wet cryocooler which is inherently a low vibration system.

%%%% of the Assuming can found using the  the decoherence rate due to the blackbody emission is given by $2\pi/\gamma_{bb}$, where $\gamma_{bb}=\frac{2\pi^4}{63}\frac{(k_BT)^6}{c^5\hbar^4\rho} Im\Bigl [\frac{\epsilon_{bb}-1}{\epsilon_{bb}+2}\Bigr ]$ and $\epsilon_{bb}$ is the dielectric constant of the particle in the blackbody emission band \cite{Chang2010}. At $T=300~$mK blackbody emission peaks around the wavelength $1~$mm. 

%%%%  one gets a motional coherence time of $\frac{2\pi}{\gamma_g}\approx 10^4~$sec, where $\gamma_g/2=\frac{8P}{\pi v \rho R}$ is the damping rate, $v=\sqrt{k_BT/m_g}$ is the velocity of the gas molecules and $m_g$ is the mass of an individual gas molecule \cite{Chang2010}. This time is significantly larger than the experimental time $t_0=10~\mu s$.In particular, to ensure a long motional coherence time of the centre-of-mass, once the desired internal temperature of the levitated particle is achieved, the trapping chamber can be evacuated to remove residual gas molecules to reach high vacuum  which guarantees a long motional coherence time. Furthermore, cryogenic environment considered is beneficial in achieving such a vacuum level.

\section{Discussion} The large spatial separation ($5~\mu$m) between the delocalized matter-wave packets that the current scheme can produce is ideal for testing wave-function collapse models such as the continuous spontaneous localization (CSL) \cite{Bassi2013}. CSL has two parameters- namely collapse rate $\Gamma_{CSL}$ and coherence length $r_{CSL}$. Assuming a successful experimental realization of the current scheme, according to CSL with $\lambda_{CSL}=1\times 10^{-17}~$s$^{-1}$, a $R=10$ nm YIG nanoparticle and a coherence time of $10~\mu$s, a collapse rate of $\Gamma=8.5\times 10^4~$Hz is predicted. Whilst Adler's version of CSL \cite{Bassi2013} predicts a collapse rate of $\Gamma=8.5\times 10^{12}~$Hz. In other words, according to the Adler version of CSL, superposition should decohere long before the time of our experiment ($10~\mu$s).

In the scale of macroscopicity $\mu_m$ \cite{NimmrichterPRL2013}, a measure of macroscopic quantumness, the experiment proposed in this article is equivalent to $16$. This is about four orders of magnitude larger than the current experimental record \cite{Kovachy2015,FrowisRevModPhys2018}. This can be boosted further by using a larger YIG nanocrystal. But, a larger nanocrystal means a greatly increased $S$ which is not ideal for an experiment (see for example, Fig. \ref{fig2}). Nevertheless, one can use a core-shell structure \cite{Neukirch2015} with a YIG core ($R=10~$nm) and the shell of a non-magnetic material such as silica of desired thickness e.g. $2~\mu$m. Of course, this will reduce $\Delta z$ significantly (see Eq. (\ref{eq2})). However, as long as the coherence time and other parameters remain unchanged, $\mu_m$ increases to $29$. More interestingly, spatial superposition of this core-shell structure can be used in the quantum gravity experiment proposed by Bose et al. \cite{BosePRL2017}. Here, one needs to ensure that the gravitational interaction between two such structures ($R\approx 2~\mu$m) dominates all other forces e.g. electric and magnetic forces \cite{BosePRL2017}. A simple comparison between the magnetic and the gravitational forces between two such microparticles shows that the gravitational attraction is three orders of magnitude stronger than the magnetic force. Here, we have used the standard magnetic dipolar interaction $\frac{6\mu_0\mu_{1}\mu_{2}}{4\pi d^4}$ and the Newtonian gravitational attraction $\frac{Gm_1m_2}{d^2}$ , where $m_1$ and $\mu_{1}$, and $m_2$ and $\mu_{2}$ are the mass and the magnetic moment of particle one and particle two, respectively. Additionally, $\mu_0$ is the magnetic permeability of free space, $G$ is the gravitational constant and $d=500~\mu$m is the distance between the two particles. To avoid Coulomb forces one can neutralize charges using electrical discharge \cite{BosePRL2017}.

\section{Conclusions} In this article we have theoretically shown that exploiting the naturally occurring spin superposition in a yttrium iron garnet nanoparticle and an appropriate magnetic field gradient, a large Schrodinger cat can be created. The spatial separation between the two arms of such a Schrodinger cat is $5~\mu$m- about $200$ times larger than the size of the particle put into the superposition. We have also shown that if successfully realized in an experiment then the current scheme will put a very strong bound on the Adler's version of wave-function collapse model. Furthermore, we have shown that a core-shell structure, a yttrium iron garnet core and a non-magnetic silica shell, in a spatial superposition can be used for testing the quantized nature of gravity.

\section*{Acknowledgement} I greatly acknowledge the financial support of UK EPSRC grant - EP/S000267/1 and the comments and suggestions that I have received from S. Bose, P. Barker and E. Chudnovsky. I also acknowledge the discussion that I had with A. Bayat and M. Toros on tunneling which  indeed initiated this article. I am indebted to J. Gosling and A. Pontin for proof reading the manuscript.


\section{References}

\begin{thebibliography}{10}
	
	\bibitem{Monroe1996}
	Monroe C, Meekhof DM, King BE, Wineland DJ.
	\newblock A {\textquotedblleft}Schr{\"o}dinger Cat{\textquotedblright}
	Superposition State of an Atom.
	\newblock Science. 1996;272(5265):1131--1136.
	
	\bibitem{Arndt1999}
	Arndt M, Nairz O, Vos-Andreae J, Keller C, Zouw GVD, Zeilinger A.
	\newblock Wave–particle duality of C60 molecules.
	\newblock Nature. 1999 October;401(6754).
	
	\bibitem{Eibenberger2013}
	Eibenberger S, Gerlich S, Arndt M, Mayor M, Txen J.
	\newblock Matterwave interference of particles selected from a molecular
	library with masses exceeding 10000 amu.
	\newblock Phys Chem Chem Phys. 2013;15:14696--14700.
	
	\bibitem{Arndt2014}
	Arndt M, Hornberger K.
	\newblock Testing the limits of quantum mechanical superpositions.
	\newblock Nat Phys. 2014;10(4):271--277.
	
	\bibitem{ZawiskyM2002}
	Zawisky M, Baron M, Loidl R, Rauch H.
	\newblock Testing the world's largest monolithic perfect crystal neutron
	interferometer.
	\newblock Nucl Instrum Methods Phys Res. 2002;481(1):406--413.
	
	\bibitem{Kovachy2015}
	Kovachy T, Asenbaum P, Overstreet C, Donnelly CA, Dickerson SM, Sugarbaker A,
	et~al.
	\newblock Quantum superposition at the half-metre scale.
	\newblock Nature. 2015;528:530--533.
	
	\bibitem{Marshall2003}
	Marshall W, Simon C, Penrose R, Bouwmeester D.
	\newblock Towards Quantum Superpositions of a Mirror.
	\newblock Phys Rev Lett. 2003 Sep;91:130401.
	
	\bibitem{RomeroPRL2012}
	Romero-Isart O, Clemente L, Navau C, Sanchez A, Cirac JI.
	\newblock Quantum Magnetomechanics with Levitating Superconducting
	Microspheres.
	\newblock Phys Rev Lett. 2012 Oct;109:147205.
	
	\bibitem{Zhang2013}
	Yin Z, Li T, Zhang X, Duan L.
	\newblock Large quantum superpositions of a levitated nanodiamond through
	spin-optomechanical coupling.
	\newblock Phys Rev A. 2013 Sep;88:033614.
	
	\bibitem{BatemanNatComm2014}
	Bateman J, Nimmrichter S, Hornberger K, Ulbricht H.
	\newblock Near-field interferometry of a free-falling nanoparticle from a
	point-like source.
	\newblock Nat Commun. 2014;5(1).
	
	\bibitem{WanPRL2016}
	Wan C, Scala M, Morley GW, Rahman ATMA, Ulbricht H, Bateman J, et~al.
	\newblock Free Nano-Object Ramsey Interferometry for Large Quantum
	Superpositions.
	\newblock Phys Rev Lett. 2016 Sep;117:143003.
	
	\bibitem{Romero_IsartNJP2017}
	Romero-Isart O.
	\newblock Coherent inflation for large quantum superpositions of levitated
	microspheres.
	\newblock N J Phys. 2017 dec;19(12):123029.
	
	\bibitem{FrowisRevModPhys2018}
	Fr\"owis F, Sekatski P, D\"ur W, Gisin N, Sangouard N.
	\newblock Macroscopic quantum states: Measures, fragility, and implementations.
	\newblock Rev Mod Phys. 2018 May;90:025004.
	
	\bibitem{RahmanATMAnishur2018}
	Rahman ATMA.
	\newblock Spatial superposition at a millimetre scale length using a levitated
	ferromagnetic nanoparticle.
	\newblock arXiv:181209948v4. 2018;.
	
	\bibitem{Bassi2013}
	Bassi A, Lochan K, Satin S, Singh TP, Ulbricht H.
	\newblock Models of wave-function collapse, underlying theories, and
	experimental tests.
	\newblock Rev Mod Phys. 2013;85(2):471--527.
	
	\bibitem{Penrose1996OGri}
	Penrose R.
	\newblock On Gravity's role in Quantum State Reduction.
	\newblock Gen Rel Gravit. 1996;28(5):581--600.
	
	\bibitem{Zurek2003}
	Zurek WH.
	\newblock Decoherence, einselection, and the quantum origins of the classical.
	\newblock Rev Mod Phys. 2003 May;75:715--775.
	
	\bibitem{Bennet2000}
	Charles HB, Divincenzo PD.
	\newblock Quantum information and computation.
	\newblock Nature. 2000;404:6775.
	
	\bibitem{QuantumMiller2008}
	Miller DAB.
	\newblock Quantum Mechanics for Scientists and Engineers.
	\newblock Cambridge University Press; 2008.
	
	\bibitem{AwschalomScience1992}
	Awschalom DD, DiVincenzo DP, Smyth JF.
	\newblock Macroscopic Quantum Effects in Nanometer-Scale Magnets.
	\newblock Science. 1992;258(5081):414--421.
	
	\bibitem{Garg1994}
	Garg A.
	\newblock Dissipation in macroscopic quantum tunneling and coherence in
	magnetic particles (invited).
	\newblock J Appl Phys. 1994;76(10):6168--6173.
	
	\bibitem{Chudnovsky_tejada_1998}
	Chudnovsky EM, Tejada J.
	\newblock Macroscopic Quantum Tunneling of the Magnetic Moment.
	\newblock Cambridge University Press; 1998.
	
	\bibitem{TejadaNanothech2001}
	Tejada J, Chudnovsky EM, del Barco E, Hernandez JM, Spiller TP.
	\newblock Magnetic qubits as hardware for quantum computers.
	\newblock Nanotechnology. 2001 may;12(2):181--186.
	
	\bibitem{HillScience2003}
	Hill S, Edwards RS, Aliaga-Alcalde N, Christou G.
	\newblock Quantum Coherence in an Exchange-Coupled Dimer of Single-Molecule
	Magnets.
	\newblock Science. 2003;302(5647):1015--1018.
	
	\bibitem{ChudnovskyPRL2000}
	Chudnovsky EM, Friedman JR.
	\newblock Macroscopic Quantum Coherence in a Magnetic Nanoparticle Above the
	Surface of a Superconductor.
	\newblock Phys Rev Lett. 2000 Dec;85:5206--5209.
	
	\bibitem{ChioleroPRB1997}
	Chiolero A, Loss D.
	\newblock Macroscopic quantum coherence in ferrimagnets.
	\newblock Phys Rev B. 1997 Jul;56:738--746.
	
	\bibitem{ChioleroPLR1998}
	Chiolero A, Loss D.
	\newblock Macroscopic Quantum Coherence in Molecular Magnets.
	\newblock Phys Rev Lett. 1998 Jan;80:169--172.
	
	\bibitem{AwschalomPRL1992}
	Awschalom DD, Smyth JF, Grinstein G, DiVincenzo DP, Loss D.
	\newblock Macroscopic quantum tunneling in magnetic proteins.
	\newblock Phys Rev Lett. 1992 May;68:3092--3095.
	
	\bibitem{GiderScience1995}
	Gider S, Awschalom D, Douglas T, Mann S, Chaparala M.
	\newblock Classical and quantum magnetic phenomena in natural and artificial
	ferritin proteins.
	\newblock Science. 1995;268(5207):77--80.
	
	\bibitem{delBarcoEPL1999}
	del Barco E, Vernier N, Hernandez JM, Tejada J, Chudnovsky EM, Molins E, et~al.
	\newblock Quantum coherence in Fe8molecular nanomagnets.
	\newblock Europhysics Letters ({EPL}). 1999 sep;47(6):722--728.
	
	\bibitem{delBarcoPRB2000}
	Barco Ed, Hernandez JM, Tejada J, Biskup N, Achey R, Rutel I, et~al.
	\newblock High-frequency resonant experiments in ${\mathrm{Fe}}_{8}$ molecular
	clusters.
	\newblock Phys Rev B. 2000;62:3018--3021.
	
	\bibitem{SchlegelPRL2008}
	Schlegel C, van Slageren J, Manoli M, Brechin EK, Dressel M.
	\newblock Direct Observation of Quantum Coherence in Single-Molecule Magnets.
	\newblock Phys Rev Lett. 2008 Oct;101:147203.
	
	\bibitem{AldaAPL2016}
	Alda I, Berthelot J, Rica RA, Quidant R.
	\newblock Trapping and manipulation of individual nanoparticles in a planar
	Paul trap.
	\newblock Appl Phys Lett. 2016;109(16):163105.
	
	\bibitem{Huillery2019}
	Huillery P, Delord T, Nicolas L, Bossche MVD, Perdriat M, Hétet G.
	\newblock Spin-mechanics with levitating ferromagnetic particles.
	\newblock arXiv:190309699. 2019;.
	
	\bibitem{VinantePRA2019}
	Vinante A, Pontin A, Rashid M, Toro\ifmmode~\check{s}\else \v{s}\fi{} M, Barker
	PF, Ulbricht H.
	\newblock Testing collapse models with levitated nanoparticles: Detection
	challenge.
	\newblock Phys Rev A. 2019 Jul;100:012119.
	
	\bibitem{VinantePRL2017}
	Vinante A, Mezzena R, Falferi P, Carlesso M, Bassi A.
	\newblock Improved Noninterferometric Test of Collapse Models Using Ultracold
	Cantilevers.
	\newblock Phys Rev Lett. 2017 Sep;119:110401.
	
	\bibitem{JohnssonSciRep2016}
	Johnsson MT, Brennen GK, Twamley J.
	\newblock Macroscopic superpositions and gravimetry with quantum
	magnetomechanics.
	\newblock Scientific Reports. 2016;6(1).
	
	\bibitem{LossPRL1992}
	Loss D, DiVincenzo DP, Grinstein G.
	\newblock Suppression of tunneling by interference in half-integer-spin
	particles.
	\newblock Phys Rev Lett. 1992 Nov;69:3232--3235.
	
	\bibitem{ChudnovskyPRB2010}
	Chudnovsky EM, Garanin DA.
	\newblock Rotational states of a nanomagnet.
	\newblock Phys Rev B. 2010 Jun;81:214423.
	
	\bibitem{WernsdorferPRL1997}
	Wernsdorfer W, Orozco EB, Hasselbach K, Benoit A, Barbara B, Demoncy N, et~al.
	\newblock Experimental Evidence of the N\'eel-Brown Model of Magnetization
	Reversal.
	\newblock Phys Rev Lett. 1997 Mar;78:1791--1794.
	
	\bibitem{PrincepQMaterials2017}
	Princep AJ, Ewings RA, Ward S, Tóth S, Dubs C, Prabhakaran D, et~al.
	\newblock The full magnon spectrum of yttrium iron garnet.
	\newblock npj Quantum Materials. 2017;2(1):1--5.
	
	\bibitem{Leuenberger2001}
	Leuenberger MN, Loss D.
	\newblock Quantum computing in molecular magnets.
	\newblock Nature. 2001;410(6830).
	
	\bibitem{Harrison2015}
	Harrison J, Hwang Y, Paydar O, Wu J, Threlkeld E, Rosenzweig J, et~al.
	\newblock High-gradient microelectromechanical system quadrupole electromagnets
	for particle beam focusing and steering.
	\newblock Phys Rev Accel Beams. 2015 Feb;18:023501.
	
	\bibitem{CHEREPANOV199381}
	Cherepanov V, Kolokolov I, L'vov V.
	\newblock The saga of YIG: Spectra, thermodynamics, interaction and relaxation
	of magnons in a complex magnet.
	\newblock Phy Rep. 1993;229(3):81 -- 144.
	
	\bibitem{Serga_2010}
	Serga AA, Chumak AV, Hillebrands B.
	\newblock {YIG} magnonics.
	\newblock J Phys D. 2010 jun;43(26):264002.
	
	\bibitem{KaplanPRL1965}
	Kaplan DE.
	\newblock Magnetostatic Mode Echo in Ferromagnetic Resonance.
	\newblock Phys Rev Lett. 1965 Feb;14:254--256.
	
	\bibitem{KaplanJAP1969}
	Kaplan DE, Hill RM, Herrmann GF.
	\newblock Amplified Ferrimagnetic Echoes.
	\newblock J Appl Phys. 1969;40(3):1164--1171.
	
	\bibitem{HueblPRL2013}
	Huebl H, Zollitsch CW, Lotze J, Hocke F, Greifenstein M, Marx A, et~al.
	\newblock High Cooperativity in Coupled Microwave Resonator Ferrimagnetic
	Insulator Hybrids.
	\newblock Phys Rev Lett. 2013 Sep;111:127003.
	
	\bibitem{RAJENDRAN2006}
	Rajendran M, Deka S, Joy PA, Bhattacharya AK.
	\newblock Size-dependent magnetic properties of nanocrystalline yttrium iron
	garnet powders.
	\newblock J Magn Magn Mater. 2006;301(1):212--219.
	
	\bibitem{MairFlaigPRB2017}
	Maier-Flaig H, Klingler S, Dubs C, Surzhenko O, Gross R, Weiler M, et~al.
	\newblock Temperature-dependent magnetic damping of yttrium iron garnet
	spheres.
	\newblock Phys Rev B. 2017 Jun;95:214423.
	
	\bibitem{CapuaAmir2017}
	Capua A, Rettner C, Yang SH, Phung T, Parkin SS.
	\newblock Ensemble-averaged Rabi oscillations in a ferromagnetic CoFeB film.
	\newblock Nat Commun. 2017;8.
	
	\bibitem{Wineland2013}
	Wineland DJ.
	\newblock Nobel Lecture: Superposition, entanglement, and raising
	Schr\"odinger's cat.
	\newblock Rev Mod Phys. 2013 Jul;85:1103--1114.
	
	\bibitem{MaminNatNano2007}
	Mamin HJ, Poggio M, Degen CL, Rugar D.
	\newblock Nuclear magnetic resonance imaging with 90-nm resolution.
	\newblock Nat Nano. 2007;2(5).
	
	\bibitem{Hinterberger_2017}
	Hinterberger A, Gerber S, Doser M.
	\newblock Superconducting shielding with Pb and Nb tubes for momentum sensitive
	measurements of neutral antimatter.
	\newblock Journal of Instrumentation. 2017 sep;12(09):T09002--T09002.
	
	\bibitem{GargPRL1993}
	Garg A.
	\newblock Dissipation by nuclear spins in macroscopic magnetization tunneling.
	\newblock Phys Rev Lett. 1993 Mar;70:1541--1544.
	
	\bibitem{SpencerPRL1959}
	Spencer EG, Lecraw RC, Clogston AM.
	\newblock Low-temperature line-width maximum in yttrium iron garnet.
	\newblock Phys Rev Lett. 1959;3(1):32--33.
	
	\bibitem{Roschmann1977}
	Röschmann P, Dötsch H.
	\newblock Properties of Magnetostatic Modes in Ferrimagnetic Spheroids.
	\newblock Phys Status Solidi B. 1977;82(1):11--57.
	
	\bibitem{KittelPR1948}
	Kittel C.
	\newblock On the Theory of Ferromagnetic Resonance Absorption.
	\newblock Phys Rev. 1948 Jan;73:155--161.
	
	\bibitem{Chang2010}
	Chang DE, Regal CA, Papp SB, Wilson DJ, Ye J, Painter O, et~al.
	\newblock Cavity opto-mechanics using an optically levitated nanosphere.
	\newblock PNAS. 2010;107(3):1005--1010.
	
	\bibitem{StillmanGregory}
	Stillman GE.
	\newblock Optoelectronics-21.
	\newblock In: Reference Data for Engineers;. p. 1--31.
	
	\bibitem{NimmrichterPRL2013}
	Nimmrichter S, Hornberger K.
	\newblock Macroscopicity of Mechanical Quantum Superposition States.
	\newblock Phys Rev Lett. 2013 Apr;110:160403.
	
	\bibitem{Neukirch2015}
	Neukirch LP, von Haartman E, Rosenholm JM, Vamivakas AN.
	\newblock Multi-dimensional single-spin nano-optomechanics with a levitated
	nanodiamond.
	\newblock Nat Photon. 2015;9:653–657.
	
	\bibitem{BosePRL2017}
	Bose S, Mazumdar A, Morley GW, Ulbricht H, Toro\ifmmode~\check{s}\else
	\v{s}\fi{} M, Paternostro M, et~al.
	\newblock Spin Entanglement Witness for Quantum Gravity.
	\newblock Phys Rev Lett. 2017 Dec;119:240401.
	
\end{thebibliography}
\end{document}